\documentclass[XeLaTex,twocolumn,epjc3]{svjour3}          

\RequirePackage[T1]{fontenc}

\smartqed  

\RequirePackage{graphicx}
\RequirePackage{mathptmx}      
\RequirePackage{flushend}
\usepackage{epstopdf}
\RequirePackage[numbers,sort&compress]{natbib}
\RequirePackage[colorlinks,citecolor=blue,urlcolor=blue,linkcolor=blue]{hyperref}

\journalname{Eur. Phys. J. C}

\begin{document}

\title{Physical viability of fluid spheres satisfying the Karmarkar condition}

\author{Ksh. Newton Singh\thanksref{e1,addr1}
        \and
        Neeraj Pant\thanksref{e2,addr2}
        \and
        M. Govender\thanksref{e3,addr3}
       }

\thankstext{e1}{e-mail: ntnphy@gmail.com}
\thankstext{e2}{e-mail: neeraj.pant@yahoo.com}
\thankstext{e3}{e-mail: megandhreng@dut.ac.za}

\institute{Department of Physics, National Defence Academy, Khadakwasla, Pune-411023, India.\label{addr1}
 \and
        Department of Mathematics, National Defence Academy, Khadakwasla, Pune-411023, India.\label{addr2}
\and
Department of Mathematics, Faculty of Applied Sciences, Durban University of Technology, Durban, South Africa.\label{addr3}}

\date{Received: date / Accepted: date}

\maketitle

\begin{abstract}
We obtain a new static model of the TOV-equation for an anisotropic fluid distribution by imposing the Karmarkar condition. In order to close the system of equations we postulate an interesting form for the $g_{rr}$ gravitational potential which allows us to solve for $g_{tt}$ metric component via the Karmarkar condition. We demonstrate that the new interior solution has well-behaved physical attributes and can be utilized to model relativistic static fluid spheres. By using observational data sets for the radii and masses for compact stars such as 4U 1538-52, LMC X-4 and PSR J1614-2230 we show that our solution describes these objects to a very good degree of accuracy. The physical plausibility of the solution depends on a parameter $c$ for a particular star. For 4U 1538-52, LMC X-4 and PSR J1614-2230 the solutions are well-behaves for $0.1574 \le c \le 0.46$, $0.1235 \le c \le 0.35$ and $0.05 \le c \le 0.13$ respectively. The behavior of the thermodynamical and physical variables of these compact objects lead us to conclude that the parameter $c$ plays an important role in determining the equation of state of the stellar material and observed that smaller values of $c$ lead to stiffer equation of states.
\end{abstract}

\section{Introduction}

The final outcome of gravitational collapse has been the focus of attention since Laplace and Michell first conceived of the idea of a black or invisible star. One of the early attempts to determine the result of continued gravitational collapse of a homogeneous dust sphere was carried out by Oppenheimer and Snyder in 1939 \cite{OppenS}. The resulting singularity remains hidden behind the trapping horizon allowing us to conclude that the final fate of collapsing homogeneous dust cloud leads to a Schwarzschild black hole. Although highly simplified the Oppenheimer-Snyder collapse model sparked an interest in seeking more general collapse scenarios \cite{joshi1,joshi2,sanjay1,harada1}. The  Cosmic Censorship Conjecture hypothesizes that any reasonable matter distribution undergoing gravitational collapse leads to the formation of a black hole i.e., singularity remains hidden behind the horizon at all times \cite{pen}. There have been a number of counterexamples to the Cosmic Censorship Conjecture with the discovery of naked singularities as possible end-states of gravitational collapse \cite{wagh1,ghosh1,ghosh2}. A natural question that arises from these investigations is how the initial static configuration (the state of the stellar fluid just before the onset of collapse) affects the outcome of gravitational collapse. To this end there have been various approaches in modeling dissipative collapse starting from an initial static configuration \cite{sharma1,shara2,bog1}. It has been shown that pressure anisotropy, shear, inclusion of charge, dimensionality of spacetime and equation of state of the initially static core affects the subsequent collapse. In a recent investigation, Naidu and Govender \cite{naidu} showed that two initially static stellar models with the same masses and radii but different pressure profiles undergoing collapse lead to very different temperature profiles, particularly during the latter stages of their evolution. Finding exact solutions of the Einstein field equations describing bounded matter distributions are important in understanding the subsequent gravitational collapse of these objects.

The Einstein field equations describing localized bodies is a system of highly nonlinear partial differential equations which are difficult to solve in general. In seeking solutions to these equations various novel ideas ranging from ad-hoc specification of the gravitational potentials, imposing an equation of state, prescribing the behavior of the density, pressure or anisotropy profiles ab initio and specifying the spacetime symmetry have been utilized. It is the very nature of the Einstein field equations which connects the curvature of spacetime to the matter content which allows one to either specify the geometry or the matter distribution to determine the behaviour of the other. The first successful attempt at modeling the interior of a spherically symmetric star was carried out by Schwarzschild in 1916 in which he considered a matter distribution with uniform density. The Schwarzschild solution is conformally flat and is characterized by isotropic pressure. Conformal flatness implies vanishing of the Weyl tensor which equates to the vanishing of tidal forces. The study of matter at ultra-high densities of the order of $10^{15} g~cm^{-3}$ indicate that the transverse and radial stresses within the stellar fluid may not be equal. Local anisotropy may drastically affect the stability of self--gravitating systems as was shown by Chan et. al \cite{chan1}. Various scenarios have been proposed to incorporate local anisotropy in stellar models some of which are pion condensation (Hartle et al. \cite{hart}), neutrino trapping at high densities \cite{Arnett} and different types of phase transitions \cite{report}. The relaxation of the pressure isotropy condition has led to an explosion of exact solutions of the Einstein field equations describing compact objects. Based on fundamental particle interactions the standard linear equation of state has been extended to include the bag constant. This equation of state has been used extensively to model compact objects with anisotropic pressure profiles as well as a non-vanishing electromagnetic field in the stellar interior. These models are well-behaved and were shown to mimic neutron stars, pulsars and strange star candidates. The quadratic equation of state has also been successfully used to model stellar interiors of compact objects such as Her X-1, RXJ 1856-37, SAX J1808.4-3658(SS1) and SAX J1808.4-3658(SS2). Utilizing curvature coordinates Herrera and Barreto derived an algorithm to generate relativistic polytropes with anisotropic pressures \cite{poly}. Motivated by the existence of dark energy, Lobo and co-workers hypothesized the existence of dark stars with an equation of state of the form $p = \alpha \rho$ where $-1 < \alpha < -1/3$ \cite{lobo1}. A more exotic form of matter distribution is the so-called Chaplygin gas and the generalized Chaplygin gas which reduce to the linear equation of state in the appropriate limit. Stable dark stars are remnants of gravitational collapse which are formed as a result of the repulsive nature of dark energy. The repulsion is sufficiently strong to halt collapse leading to the formation of stable stars free of any singularity \cite{p1,p2}.

Higher order gravity theories have been fruitful in producing models of compact stellar objects. Various authors have shown that modifications to 4-D classical Einstein gravity feature in the thermodynamical properties of the stellar fluid \cite{sud1,sud2,abbas}. The braneworld scenario provides a natural mechanism for the existence of anisotropic pressure within the stellar fluid \cite{brig}. In addition, it was shown that in the Randall-Sundrum II type braneworld, the exterior spacetime of spherical star is filled with radiative-type stresses induced by 5-dimensional graviton effects and is not necessarily the vacuum Schwarzschild solution as in the 4-D case \cite{roy1}. Recently, Dadhich et. al \cite{iucaa} have shown that within the framework of pure Lovelock gravity there cannot exist self-gravitating bounded distributions  $d =2N + 1$ dimensions. This is to say that there is no finite radius for which the pressure vanishes \cite{iucaa}. The transition from classical 4-D gravity to higher dimensional gravity theories has sparked immense interest in studying phenomenological processes which reside in extra dimensions. One of the main proponents of these investigations is Dadhich and his collaborators who proved the universality of the Schwarzschild constant density sphere, ie. it was shown that this solution carries over to Einstein-Gauss-Bonnet gravity and Lovelock gravity \cite{s1,s2}.

It is widely believed that the four fundamental interactions in Nature were once a manifestation of a single, unified force. Furthermore, the dimensionality of spacetime could have evolved in such a manner so as to reveal four dimensions which we observe today. Kaluzua-Klein theories have shown that the electromagnetic interaction manifests naturally in 5-dimensional spacetime. These observations of physical phenomena transcending the dimensionality of spacetime have generated widespread interest in embedding our standard four-dimensional spacetime into higher dimensional spacetimes \cite{higher}. It is well-known that any  pseudo-Riemannian manifold, $({\cal V}_n)^{-}$ with dimensionality $n$ may be locally embedded into a pseudo-Euclidean space, $({\cal V}_m)^{+}$ of dimension $m = \frac{n(n + 1)}{2}$. It follows that the embedding class of $({\cal V}_n)^{-} \leq m - n = \frac{n(n - 1)}{2}$. For the relativistic 4-D spacetime $({\cal V}_4)^{-}$, the embedding class is 6. A recent and popular approach in deriving exact solutions of the Einstein field equations describing compact stars is to make use of the Karmarkar condition \cite{k1,k3,k7,kn1,kn2,kns1,kns2,kns3,kns4}. The necessary and sufficient condition for a spherically symmetric spacetime to be of embedding class I was first derived by Karmarkar \cite{kar48}. It is a mathematical simplification which reduces the problem of obtaining exact solutions to a single-generating function. The approach is to choose one of the gravitational potentials on physical grounds and to then integrate the Karmarkar condition to fully specify the gravitational behavior of the model. In this paper we utilize the Karmarkar condition to derive solutions which describe compact objects in general relativity. We subject our solutions to rigorous physical tests which ensure that they do describe physically observable objects in the universe.

\section{Einstein field equations for anisotropic fluid distributions}

The interior of the super-dense star is assumed to be described by the line element
\begin{equation}
ds^2 = -e^{\nu(r)}dt^2 + e^{\lambda(r)}dr^2 + r^2(d\theta^2 + \sin^2{\theta}d\phi^2) \label{metric}
\end{equation}
where the gravitational potentials $\nu(r)$ and $\lambda(r)$ are yet to be specified.
The Einstein field equations describing an anisotropic fluid distribution are given as (in the unit $G=c=1$)
\begin{equation} \label{efe}
-8\pi T^\mu_\xi = \mathcal{R}^\mu_\xi-{1\over 2}\mathcal{R}~g^\mu_\xi
\end{equation}
where
\begin{eqnarray}
T^\mu_\xi & = & \rho v^\mu v_\xi + p_r \chi_\xi \chi^\mu + p_t(v^\mu v_\xi -\chi_\xi \chi^\mu-g^\mu_\xi) ~,
\end{eqnarray}
is the energy-momentum tensor, $\mathcal{R}^\mu_\xi$  is the Ricci tensor, $\mathcal{R}$ represents the scalar curvature, $p_r$ and $p_t$ denote radial and transverse pressures respectively, $\rho$ is the density of the fluid distribution , $v^\mu$  the four velocity and $\chi^\mu$  is the unit space-like vector in the radial direction.

The Einstein field equations (\ref{efe}) for the line element (\ref{metric}) are
\begin{eqnarray}\label{g3}
8\pi \rho(r) &=&
\frac{1 - e^{-\lambda}}{r^2} + \frac{\lambda'e^{-\lambda}}{r} \label{g3a} \\ \nonumber \\
8\pi p_r(r) &=&  \frac{\nu' e^{-\lambda}}{r} - \frac{1 - e^{-\lambda}}{r^2} \label{g3b}
\end{eqnarray}
\begin{eqnarray}
8\pi p_t(r) &=& \frac{e^{-\lambda}}{4}\left(2\nu'' + {\nu'}^2  - \nu'\lambda' + \frac{2\nu'}{r}-\frac{2\lambda'}{r}\right) \label{g3c}
\end{eqnarray}

where primes denote differentiation with respect to the radial coordinate $r$. In generating the above field equations we have utilized geometrized units where the coupling constant and the speed of light are taken to be unity. Using Eqs. (\ref{g3b}) and (\ref{g3c}) we obtain the anisotropic parameter

\begin{eqnarray}
\Delta(r) &=& 8\pi (p_t-p_r) \nonumber \\
&=& e^{-\lambda}\left[{\nu'' \over 2}-{\lambda' \nu' \over 4}+{\nu'^2 \over 4}-{\nu'+\lambda' \over 2r}+{e^\lambda-1 \over r^2}\right] \label{del}
\end{eqnarray} which vanishes in the case of isotropic pressure.

Eisenhart \cite{eis} has mentioned that for any Riemannian space to be class I, a necessary and sufficient condition is that there exist a second-order symmetric tensor $b_{\mu \alpha}$ satisfying the following equations:
\begin{eqnarray}
\mathcal{R}_{\mu \nu \alpha \beta} &=& \epsilon (b_{\mu \alpha}b_{\nu \beta}-b_{\mu \beta}b_{\nu \alpha}) \label{ac}\\
0 &=& b_{\mu \nu ;\alpha}-b_{\mu \alpha ;\nu}  
\end{eqnarray}

where $\epsilon= \pm 1$ ($+$ when the normal to the manifold is space-like or $-$ when the normal to the manifold is time-like) and `(;)' represents covariant differentiation.

For the line element (\ref{metric}), the non-zero components of the Riemann curvature tensor are given below:
\begin{eqnarray}
\mathcal{R}_{1414} &=& -e^\nu \left({\nu'' \over 2}+{\nu'^2 \over 4}-{\lambda' \nu' \over 4}\right) \label{re1}\\
\mathcal{R}_{2323} &=& -e^\lambda r^2 \sin ^2\theta~ (e^\lambda-1) \label{a1}\\
\mathcal{R}_{1334} &=& \mathcal{R}_{1224}~\sin ^2\theta = 0 \label{ab}\\
\mathcal{R}_{1212} &=& {1\over 2}r \lambda'\\
\mathcal{R}_{3434} &=& -{1\over 2}r\sin ^2\theta~ \nu' e^{\nu-\lambda}\label{re2}
\end{eqnarray}

The non-zero components of the tensor $b_{\mu \alpha}$ corresponding to (\ref{metric}) are $b_{11},~b_{22},~b_{33}, ~b_{44}$ and $b_{14}$ with $b_{33}=b_{22}\sin^2 \theta$. With these components, (\ref{ac}) reduces to 
\begin{equation}
\mathcal{R}_{1414}={\mathcal{R}_{1212}\mathcal{R}_{3434}+\mathcal{R}_{1224}\mathcal{R}_{1334} \over \mathcal{R}_{2323}}\label{con}
\end{equation}
which is known as the Karmarkar condition \cite{kar48} in literature.

Using (\ref{re1})-(\ref{re2}) in (\ref{con}) leads to the following differential equation
\begin{equation}
{\lambda' e^\lambda \over e^\lambda-1} = {2\nu'' \over \nu'}+\nu' \label{dif1}
\end{equation}

which can be easily integrated to give a relationship between $\nu(r)$ and $\lambda(r)$ as
\begin{equation}
e^\lambda = 1 + \frac{K{\nu'}^2e^\nu}{4}\label{nu1}
\end{equation}
where $K$ is constant of integration.

By using (\ref{nu1}) we can rewrite (\ref{del}) as
\begin{equation}
\Delta(r) = {\nu' \over 4e^\lambda}\left[{2\over r}-{\lambda' \over e^\lambda-1}\right]~\left[{\nu' e^\nu \over 2rB^2}-1\right],~~~~B={1\over \sqrt{K}} \label{del1}
\end{equation}

However, Pandey \& Sharma \cite{pandey} argued that satisfying Karmarkar condition alone is insufficient for a spherically symmetric spacetime be class I. As an example, they presented the following spacetime
\begin{eqnarray}
ds^2=-e^\nu dt^2+dr^2+r^2(d\theta^2+\sin^2 \theta d\phi^2) \label{pan}
\end{eqnarray}
which does satisfy (\ref{con}). This spacetime (\ref{pan}) has $e^\lambda=1$ that imply to $\mathcal{R}_{2323}= 0$ from (\ref{a1}). $e^\lambda=1$ or $\mathcal{R}_{2323}= 0$ also implies (\ref{pan}) is spatially flat.

Now the non-zero components of curvature tensor for (\ref{pan}) are $\mathcal{R}_{1414},~\mathcal{R}_{2424}$ and $\mathcal{R}_{3434}$ only.  Using these components, (\ref{ac}) imply inconsistent equations:
\begin{eqnarray}
b_{22}b_{33}=0,~b_{24}b_{33}=0,~b_{22}b_{44}-b_{24}^2 \neq 0,~b_{33}b_{44}\neq 0.
\end{eqnarray}  
Therefore, the spacetime given in (\ref{pan}) does satisfy Karmarkar condition but fails to satisfy (\ref{ac}) i.e. (\ref{pan}) is not a class I spacetime due $\mathcal{R}_{2323}= 0$. Hence, any symmetric spacetime are called class I if they satisfy Karmarkar condition and Pandey-Sharma condition ($\mathcal{R}_{2323}\ne 0$) simultaneously. Since the condition $\mathcal{R}_{2323}= 0$ or equivalently $e^\lambda = 1$ gives the spacetime (\ref{pan}), which in fact describes a perfect fluid sphere with zero density \cite{pandey}. Hence, in order to describe perfect fluid with non-vanishing density we require $\mathcal{R}_{2323} \neq 0$. It is also important to note that all the spherically symmetric spacetimes are in general class II unless they simultaneously satisfy Karmarkar and Pandey-Sharma conditions.

\section{Isotropic Class I solutions }
For isotropy in pressure, the anisotropy factor $\Delta = 0$. Assuming that $\nu'(r) \ne 0$, we will get from (\ref{del1}) either
\begin{eqnarray}
\left[{2\over r}-{\lambda' \over e^\lambda-1}\right] &=& 0~~~~\mbox{or} \label{sch}\\
\left[{\nu' e^\nu \over 2rB^2}-1\right] &=& 0 \label{kohl}
\end{eqnarray}
or both. The first condition (\ref{sch}) leads to Schwarzschild's constant density model \cite{schw} and second condition (\ref{kohl}) leads to Kohler-Chao solution \cite{kc}.

\subsection{Schwarzschild interior solution}
Integration of (\ref{sch}) yields
\begin{equation}
e^{-\lambda}=1-cr^2 \label{sc1}
\end{equation}
Using (\ref{sc1}) in (\ref{nu1}), we obtain
\begin{equation}
e^\nu=\Big(A-{B \over \sqrt{c}} \sqrt{1-cr^2}\Big)^2
\end{equation}
The above solution is the well-known interior Schwarzschild model which describes an incompressible, static sphere with uniform density.
For completeness we present the physical quantities of this solution as determined from (\ref{g3a}) and (\ref{g3b})
\begin{eqnarray}
\rho(r) &=& {3c \over 8\pi}\\
P(r) &=& {c \over 8\pi} \left(\frac{2 B \sqrt{1-c r^2}}{A \sqrt{c}-B \sqrt{1-c r^2}}-1\right)\\
{P(r) \over \rho(r)} &=& \frac{1}{3} \left(\frac{2 B \sqrt{1-c r^2}}{A \sqrt{c}-B \sqrt{1-c r^2}}-1\right)
\end{eqnarray}

The interior Schwarzschild solution has been extensively studied by various authors including Schwarzschild himself \cite{schw}. This solution serves as a toy model for self-gravitating bounded configurations. One of its main shortcomings is the fact that it leads to infinite speed of sound within the interior of the sphere.

\subsection{Kohler-Chao solution: a cosmological solution}

Integrating (\ref{kohl})  we obtain
\begin{equation}
e^{\nu}=A+B r^2 \label{koh1}
\end{equation}
and using (\ref{koh1}) in (\ref{nu1}) yields
\begin{equation}
e^{\lambda}={A+2Br^2 \over A+Br^2} \label{kohe}
\end{equation}
which is the Kohler-Chao-Tikekar solution \cite{kc,tik}.

The corresponding expressions for density,  pressure and equation of state parameter can be written as
\begin{eqnarray}
8\pi \rho(r) &=& {B(3A+2Br^2) \over (A+2Br^2)^2}\\
8\pi P(r) &=& {B \over A+2Br^2} \label{per}\\
{P(r) \over \rho(r)} &=& \frac{A+2 B r^2}{3 A+2 B r^2}
\end{eqnarray}
with $B > 0$. However, we can see clearly from (\ref{per}) that the pressure at the surface of any configuration can't be zero for a finite boundary unless the boundary itself is infinite. This property of infinite boundary does have the property of a cosmological solution. The same discussion is also given in Maurya et al. \cite{maurya}.

Maurya et al. \cite{maurya} comments on the charged isotropic solutions of embedding Class I i.e. Schwarzschild interior and Kohler-Chao solutions. If the charge vanishes in these two solutions, then the remaining neutral counterpart will only be either the Schwarzschild interior solution or the Kohler-Chao solution, otherwise either the charge cannot be zero or the surviving space-time metric will become flat.

It is well-known that an isotropic spherically symmetric conformally flat metric is necessarily to be a class I solution. However, whether the converse holds good i.e. is a spherically symmetric class I solution representing isotropic fluid sphere necessarily conformally flat? This was resolved by Tikekar \cite{tik} concluding that ``it is not necessary a spherically symmetric class I solution representing isotropic fluid sphere be conformally flat'' and he gave an example by rediscovering the Kohler-Chao solution. Here what we want to stress is that the conformally flat solution i.e. the Schwarzschild interior solution is the only isotropic class I solution that can represent a bounded stellar configuration. However, the conformally non-flat solution i.e. the Kohler-Chao solution cannot describe a finite bounded configuration although it can qualify as a cosmological solution.

It is well-known that all the non-vanishing components of Weyl tensor are proportional to
\begin{eqnarray}
W &=& \frac{r^3e^{-\lambda}}{6}\left[\frac{e^\lambda}{r^2} - \frac{1}{r^2} + \frac{\nu' \lambda'}{4} - \frac{\nu'^2}{4} - \frac{\nu''}{2}  +\frac{\nu'-\lambda'}{2r}\right]
\end{eqnarray}
Here the Schwarzschild interior solution yields a vanishing Weyl tensor ($W=0$) postulating that it is a conformally flat space. However, the Kohler-Chao solution yields non-vanishing Weyl tensor where
\begin{equation}
W = \frac{2B^2r^5}{3(A + 2Br^2)^2}
\end{equation}
implying that the Kohler-Chao solution is not conformally flat. In general, the Karmarkar condition and pressure isotropy does not imply conformal flatness. However, the converse is true: Conformally flat, perfect fluid spheres obey the Karmarkar condition.

\section{Generating a new family of embedding class I models}

We now seek relativistic stellar models which satisfy the Karmarkar condition. In light of our findings in the previous section we relax the condition of pressure isotropy. This implies that the radial and tangential stresses are unequal throughout the fluid distribution. It is well-known that pressure anisotropy plays an important role during dissipative collapse. In a recent study by Govender et al. \cite{rob} it has been shown that the dynamics of a collapsing core is closely related to the radial pressure and energy density of the stellar fluid. By assuming a linear equation of state for the initial static configuration of the form $p_r = \alpha \rho - \beta$ where $\alpha$ and $\beta$ are constants, they demonstrated that the subsequent collapse is sensitive to the interplay between the radial pressure and energy density. They also demonstrated that the equation of state parameter, $\alpha$ influences the behaviour of the temperature profile of the collapsing body.

We now proceed to obtain a family of solutions which describe anisotropic matter configurations obeying the Karmarkar condition. In order to completely specify the gravitational behavior of our model we assume
\begin{equation}
e^\lambda = a r^2 \sin ^2\left(b r^2+c\right)+1 \label{elam}
\end{equation}
where  $a$, $b$ and $c$ are constants which are determined from the boundary conditions. The sinusoidal behavior of the gravitational potential has been widely used in various contexts in both cosmology and astrophysics. Dadhich and Raychaudhuri demonstrated that it was possible to obtain an oscillating cosmological model without Big Bang singularity \cite{ray1}. An interesting feature of this model is that it allows for the prediction of blue-shifts without violating the basic postulates of general relativity. In modeling dissipative gravitational collapse of a spherically symmetric star in which the Weyl stresses vanish, Maharaj and Govender \cite{meg1} showed that the solution of the boundary condition admits oscillatory solutions. The extension from 4-D to 5-D gravity of the Finch and Skea stellar model leads to sinusoidal behavior of the gravitational potentials \cite{sud1}.

Using the metric potential (\ref{elam}) in (\ref{nu1}), we get
\begin{eqnarray}
e^\nu = \left[A-\frac{\sqrt{a} B }{2 b}~\cos \left(b r^2+c\right)\right]^2\label{enu}
\end{eqnarray}

Using (\ref{elam}) and (\ref{enu}), we can rewrite the expression of density, $p_r$, $\Delta$ and $p_t$ as
\begin{eqnarray}
8\pi \rho(r) & = & {a \over \left[a r^2 \sin ^2\left(b r^2+c\right)+1\right]^2} ~\bigg[a r^2 \sin ^4\left(b r^2+c\right) \nonumber \\
&& +3 \sin ^2\left(b r^2+c\right)+2 b r^2 \sin \left\{2 \left(b r^2+c\right)\right\}\bigg]\\
8\pi p_r(r) & = & \frac{\sqrt{a} ~\sin \left(b r^2+c\right)}{2 \left[a r^2 \sin ^2\left(b r^2+c\right)+1\right]} \times \\
&& \frac{4 \sqrt{a} A b \sin \left(b r^2+c\right)-a B \sin \left\{2 \left(b r^2+c\right)\right\}-8 b B}{\sqrt{a} B  \cos \left(b r^2+c\right)-2 A b} \nonumber \\
\\
\Delta(r) &=& \frac{r ~\csc ^4\left(b r^2+c\right)}{4 \left[a r^2+\csc ^2\left(b r^2+c\right)\right]^2} \times \nonumber \\
&& \frac{a \cos \left\{2 \left(b r^2+c\right)\right\}-a+4 b \cot \left(b r^2+c\right)}{2 A b-\sqrt{a} B \cos \left(b r^2+c\right)}\times \nonumber\\
&& \bigg[2 a A b r \cos \left\{2 \left(b r^2+c\right)\right\}-2 a A b r +4 \sqrt{a} b B r  \nonumber\\
&& \sin \left(b r^2+c\right)+ a^{3/2} B r \sin \left(b r^2+c\right) \nonumber\\
&& \sin \left\{2 \left(b r^2+c\right)\right\} \bigg]\\
8\pi p_t(r) &=& 8\pi p_r(r)+\Delta(r)
\end{eqnarray}

\section{Properties of the new model}

The central values of pressures and density are given by
\begin{eqnarray}
8\pi p_{rc} & = & 8\pi p_{tc} \nonumber \\
&  = & \frac{\sqrt{a}~ \sin c \big(4 \sqrt{a} A b \sin c-a B \sin (2 c)-8 b B\big)}{2 \big(\sqrt{a} B \cos c-2 A b\big)}\label{pc}\\
8\pi \rho_{c} & = & 3 a \sin ^2c
\end{eqnarray}

To satisfy Zeldovich's condition at the interior, $p_{rc}/\rho_c$ at center must be $\le 1$. Therefore,
\begin{eqnarray}
\frac{4 \sqrt{a} A b \sin c-a B \sin (2 c)-8 b B}{3 \sqrt{a} \sin c\big[2 \sqrt{a} B \cos c-4 A b\big]} \le 1 \label{zel}
\end{eqnarray}

On using (\ref{pc}) and (\ref{zel}) we generate a constraint on $B/A$ given as
\begin{equation}
{8b+a \sin (2c) \over 4\sqrt{a} Ab \sin c} <{A \over B} \le {a \sin (2c) +2b \over 4\sqrt{a} b \sin c}
\end{equation}

\section{Matching of physical boundary conditions}

The exterior spacetime of our static model is the vacuum Schwarzschild solution given by
\begin{eqnarray}
ds^2 &=& \left(1-{2M\over r}\right) dt^2-\left(1-{2M\over r}\right)^{-1}dr^2 \nonumber\\
& & -r^2(d\theta^2+\sin^2 \theta d\phi^2) \label{ext}
\end{eqnarray}

\begin{table*}
\caption{Parameters of four well-known compact stars that give masses and radii compatible with observational data.}
\label{tab}
\begin{tabular*}{\textwidth}{@{\extracolsep{\fill}}lrrrrrrrrl@{}}
\hline
\multicolumn{1}{c}{$a~(km^{-2})$} & \multicolumn{1}{c}{$b~(km^{-2})$} & \multicolumn{1}{c}{$A$} & \multicolumn{1}{c}{$B$ (km$^{-1}$)} & \multicolumn{1}{c}{$c$} & \multicolumn{1}{c}{$R$ (km)} & \multicolumn{1}{c}{$M/M_\odot$} & \multicolumn{1}{c}{$u=2M/r_b$} & \multicolumn{1}{c}{$z_s$} & Object \\
\hline
0.1217 & 0.00025 & 21.343 & 0.0299 & 0.18 & 7.866 & 0.87 & 0.22 & 0.133 & 4U1608-52 \\
0.1826 & 0.00020 & 32.650 & 0.0302 & 0.15 & 8.300 & 1.04 & 0.25 & 0.154 & LMC X-4  \\
0.6123 & 0.00010 & 128.73 & 0.0329 & 0.10 & 9.690 & 1.97 & 0.41 & 0.299 & PSR J1614-2230 \\
\hline
\end{tabular*}
\end{table*}

By matching the first and second fundamental forms the interior solution (\ref{metric}) and exterior solution (\ref{ext}) at the boundary $r=R$ (Darmois-Israel junction conditions) we obtain
\begin{eqnarray}
e^{\nu_b} &=& 1-{2M \over R} = \left[A-\frac{\sqrt{a} B }{2 b}~\cos \left(b R^2+c\right)\right]^2\label{bou1}\\
e^{-\lambda_b} &=& 1-{2M \over R} = \Big[a R^2 \sin ^2\left(b R^2+c\right)+1\Big]^{-1} \label{bou2}\\
p_r(R) &=& 0 \label{bou3}
\end{eqnarray}

Using the boundary condition (\ref{bou1}-\ref{bou3}),  we get
\begin{eqnarray}
B &=& \frac{4 \sqrt{a} A b \sin \left(b R^2+c\right)}{a \sin \left\{2 \left(b R^2+c\right)\right\}+8 b} \label{b}\\
A & = & \frac{\sqrt{1-2 M/R} \Big[a \sin \left\{2 \left(b R^2+c\right)\right\}+8 b\Big]}{a \sin \left\{2 \left(b R^2+c\right)\right\}-a \sin \left(2 b R^2+2 c\right)+8 b} \\
a &=& \bigg[\frac{1}{1-2 M/R}-1\bigg]~{1 \over R^2 \sin ^2\left(b R^2+c\right)} \label{a}
\end{eqnarray}
and we have chosen $b,~c,~M$ and $R$ as free parameters and the rest of the constants $a,~A$ and $B$ are determined from the Eqs. (\ref{b}-\ref{a}).

The gravitational red-shift of the stellar system is given by
\begin{eqnarray}
Z(r) & = & \left[A-\frac{\sqrt{a} B }{2 b} ~\cos \left(b r^2+c\right)\right]^{-1}-1
\end{eqnarray}

The mass-radius relation and compactness parameter of the solution can be determined using the equation given below:
\begin{eqnarray}
m(r) & = & 4\pi \int_0^r \rho r^2 dr = \frac{a r^3 \sin ^2\left(b r^2+c\right)}{a r^2+2-a r^2 \cos \left(2 b r^2+2 c\right)}\\
u(r) &=& {2m(r) \over r} = \frac{2a r^2 \sin ^2\left(b r^2+c\right)}{a r^2+2-a r^2 \cos \left(2 b r^2+2 c\right)}
\end{eqnarray}

\section{Equilibrium and stability conditions}

\subsection{Condition for equilibrium}

For a stellar system in equilibrium under different forces, the generalized Tolman-Oppenheimer-Volkoff (TOV) equation must be satisfied \citep{ponce} i.e.
\begin{eqnarray}
{2\Delta \over r} = {dp_r \over dr}+{M_g(\rho+p_r) \over r^2}~e^{(\lambda-\nu)/2}  \label{tove}
\end{eqnarray}
where $M_g(r)$ is the effective gravitational mass contained within a sphere of radius $r$ and is defined by the Tolman-Whittaker formula viz.,
\begin{eqnarray}
M_g(r) &=& 4 \pi \int_0^r \big(T^t_t-T^r_r-T^\theta_\theta-T^\phi_\phi \big) r^2 ~e^{(\nu+\lambda)/2}dr \label{mrho1}
\end{eqnarray}

For the Eqs. (\ref{g3a})-(\ref{g3c}), the above Eq. (\ref{mrho1}) reduces to
\begin{eqnarray}
M_g(r) &=& {1\over 2}~r^2 \nu' ~e^{(\nu-\lambda)/2}
\end{eqnarray}

Equation (\ref{tove}) can be written in terms of balanced force equation due to anisotropy ($F_a$), gravity ($F_g$) and hydrostatic ($F_h$) i.e.
\begin{equation}
F_g+F_h+F_a=0 \label{forc}
\end{equation}

Here
\begin{eqnarray}
F_g & = & -{M_g(\rho+p_r) \over r^2}~e^{(\lambda-\nu)/2}\\
F_h &=& -{dp_r \over dr}\\
F_a &=& {2\Delta \over r}\\
\end{eqnarray}
The TOV equation (\ref{forc}) can be represented graphically showing the interplay amongst $F_g$, $F_h$ and $F_a$ required to bring about equilibrium as evidenced in Fig. \ref{tov}.

\subsection{Relativistic adiabatic index and stability}

For a relativistic anisotropic sphere the stability is related to the adiabatic index $\Gamma$, the ratio of two specific heats, defined by \cite{chan1},
\begin{equation}
\Gamma=\frac{\rho+p_r}{p_r}\frac{dp_r}{d\rho}.
\end{equation}

Now $\Gamma>4/3$ gives the condition for the stability of a Newtonian sphere and $\Gamma =4/3$ being the condition for a neutral equilibrium proposed by \cite{bondi64}. This condition changes for a relativistic isotropic sphere due to the regenerative effect of pressure, which renders the sphere more unstable. For an anisotropic general relativistic sphere the situation becomes more complicated, because the stability will depend on the type of anisotropy. For an anisotropic relativistic sphere the stability condition is given by \cite{chan1},

\begin{equation}
\Gamma>\frac{4}{3}+\left[\frac{4}{3}\frac{(p_{t0}-p_{r0})}{|p_{r0}^\prime|r}+\frac{8\pi}{3}\frac{\rho_0p_{r0}}{|p_{r0}^\prime|}r\right]_{max},
\end{equation}
where, $p_{r0}$, $p_{t0}$, and $\rho_0$ are the initial radial, tangential, and energy density in static equilibrium satisfying (\ref{tove}). The first and last term inside the square brackets represent the anisotropic and relativistic corrections respectively and both the quantities are positive which increase the unstable range of $\Gamma$ \citep{her79,chan1}.

\subsection{Causality and stability condition}

\begin{figure} 
\begin{minipage}{\columnwidth}
\centering
\includegraphics[scale=0.6]{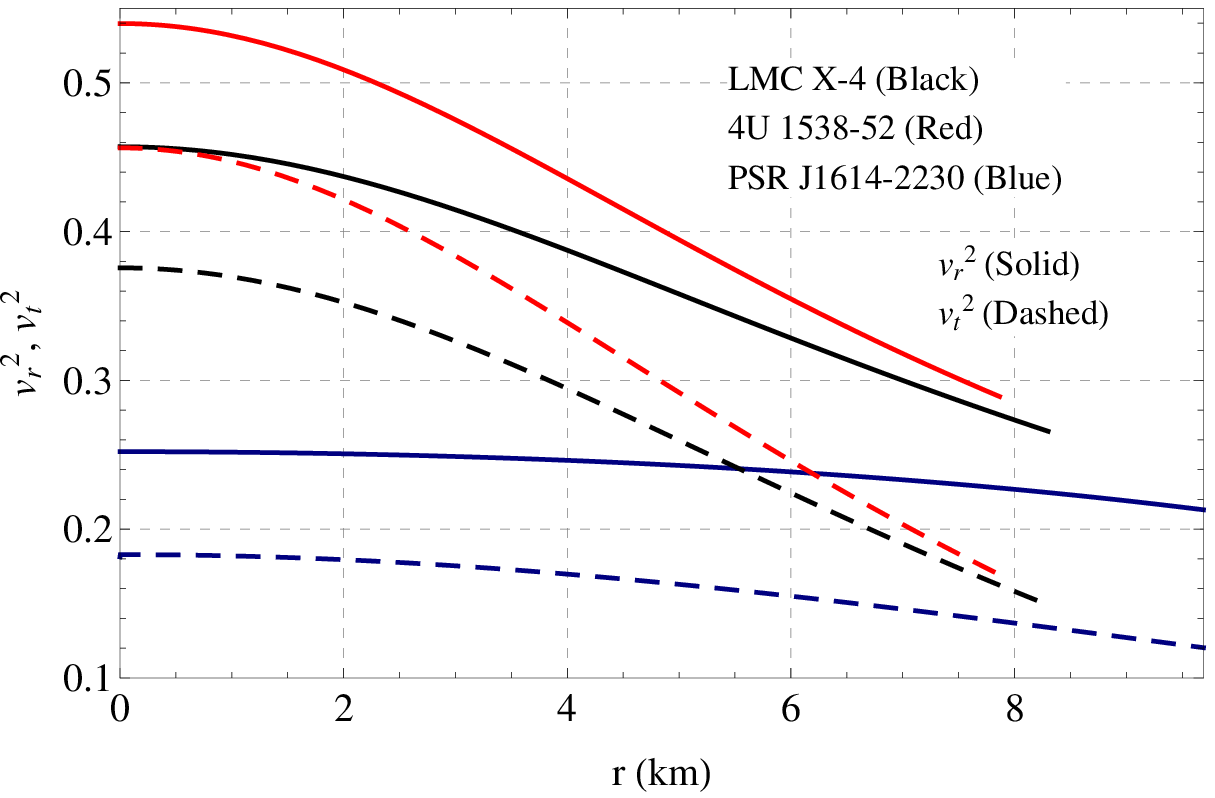}
\end{minipage}
\caption{Variation of $v_r^2$ and $v_t^2$ with radial coordinate $r$ for 4U1538-52, LMC X-4 and PSR J1614-2230 with their respective parameters given in Table \ref{tab}.}
\label{sound}
\end{figure}

The radial and tangential speeds of sound of our compact star model are given by,
\begin{equation}
v_r^2=\frac{dp_r}{d\rho}={dp_r/dr \over d\rho/dr},~~~v_t ^2=\frac{dp_t}{d\rho}={dp_t/dr \over d\rho/dr}
\end{equation}
The profile of $v_r^2$ and $v_t^2$ are given in Fig. \ref{sound} which indicates that both the radial and transverse velocity satisfy the causality conditions, i.e., both $v_r^2,\,v_t^2$ are less than $1$ and monotonic decreasing function of $r.$

\begin{figure}[t] 
\begin{minipage}{\columnwidth}
\centering
\includegraphics[scale=0.6]{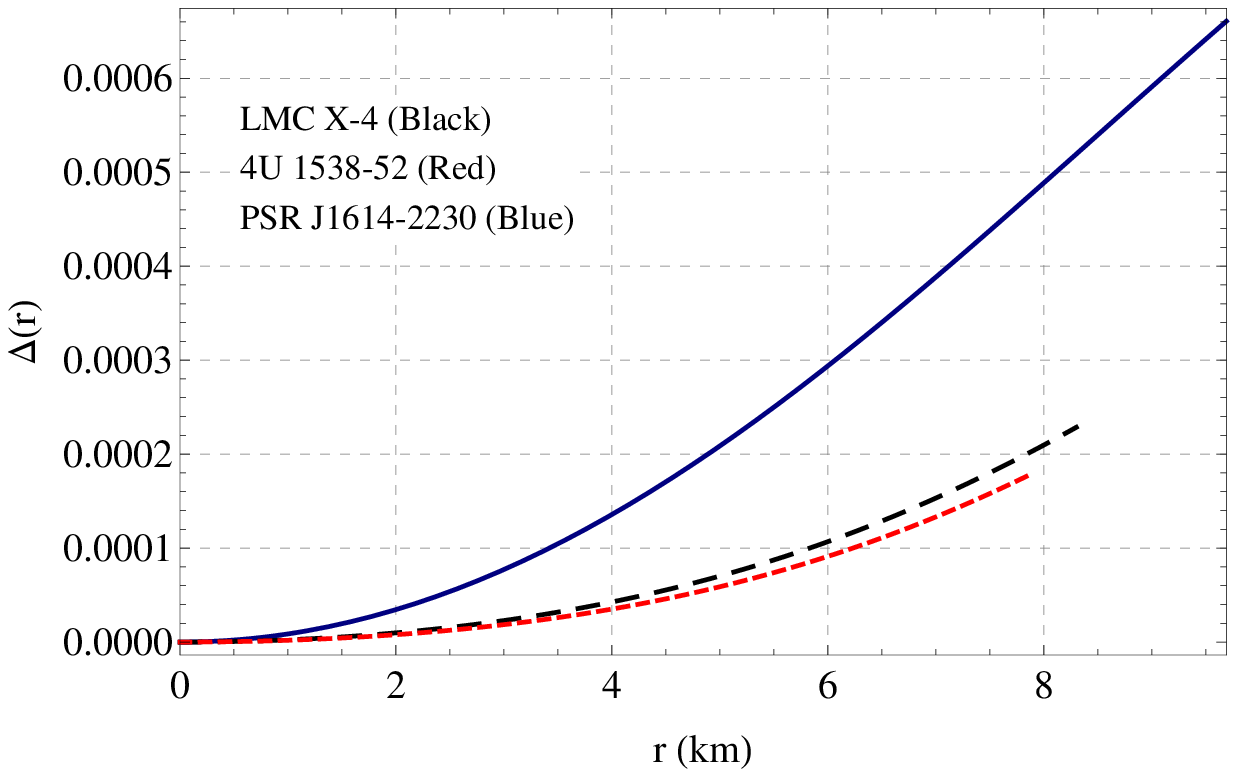}
\end{minipage}
\caption{Variation of anisotropy (km$^{-2}$) with radial coordinate $r$ for 4U1538-52, LMC X-4 and PSR J1614-2230 with their respective parameters given in Table \ref{tab}.}
\label{aniso}
\end{figure}

\begin{figure} 
\begin{minipage}{\columnwidth}
\centering
\includegraphics[scale=0.6]{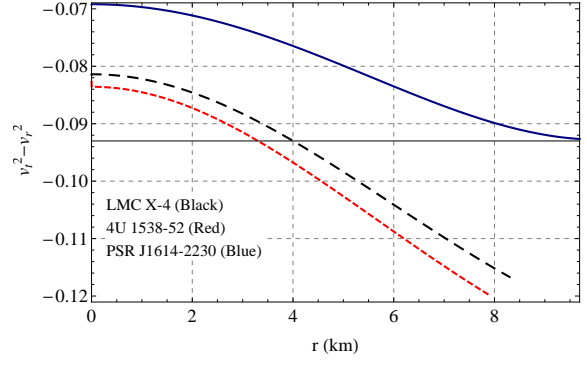}
\end{minipage}
\caption{Variation of stability factor $v_t^2-v_r^2$ ~ with radial coordinate $r$ for 4U1538-52, LMC X-4 and PSR J1614-2230 with their respective parameters given in Table \ref{tab}.}
\label{stab}
\end{figure}

\begin{figure} 
\begin{minipage}{\columnwidth}
\centering
\includegraphics[scale=0.6]{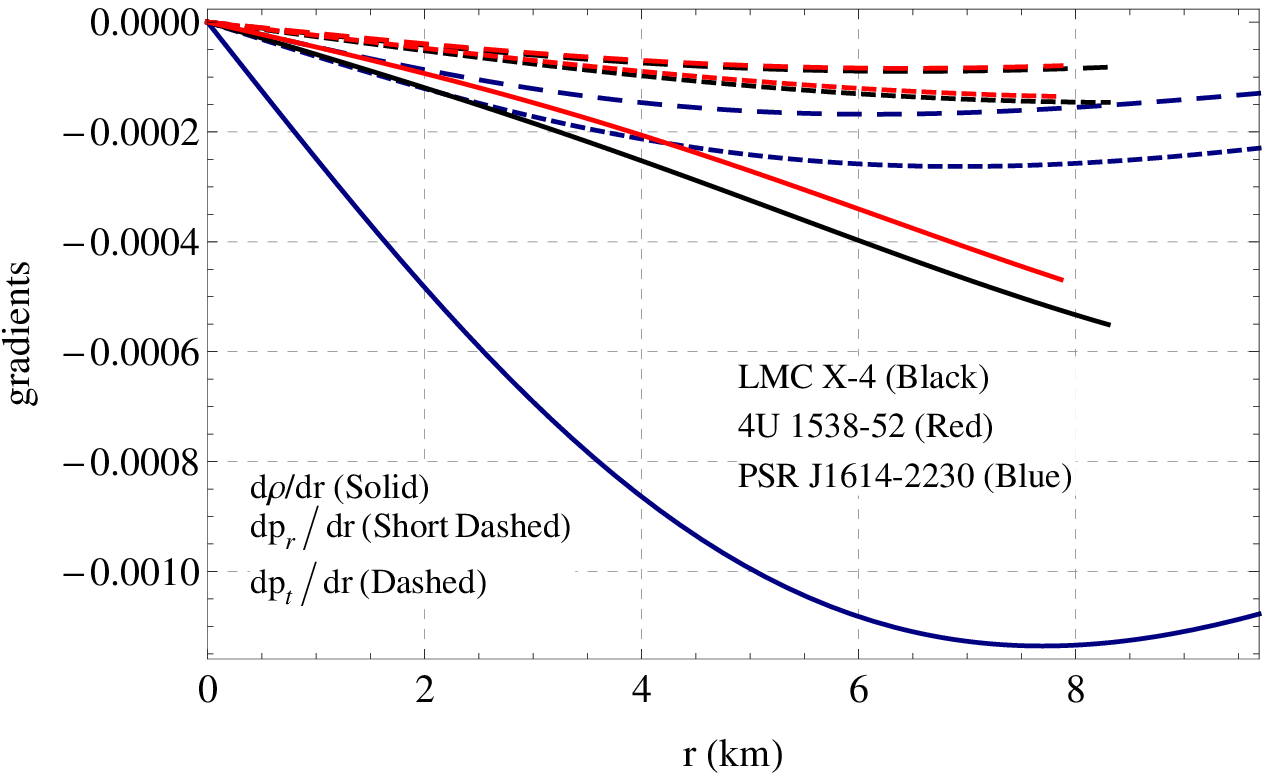}
\end{minipage}
\caption{Variation of $d\rho/dr,~dp_r/dr$ and $dp_t/dr$ (km$^{-1}$) with radial coordinate $r$ for 4U1538-52, LMC X-4 and PSR J1614-2230 with their respective parameters given in Table \ref{tab}.}
\label{grad}
\end{figure}

The stability of anisotropic stars under the radial perturbations  is studied by using the concept of \cite{her92} known as Hererra's ``cracking" method. Using the concept of cracking, \cite{Abreu} showed that the region of the anisotropic fluid sphere where $-1\leq v_{t}^{2}-v_{r}^{2}\leq0$ is potentially stable but the region where $0<v_{t}^{2}-v_{r}^{2}\leq 1$ is potentially unstable.

\begin{eqnarray}\label{crak}
  \frac{dp_t}{d\rho} &=& \frac{dp_r}{d\rho}+\frac{d\Delta}{d\rho} = \frac{dp_r}{d\rho}+\frac{d\Delta /dr}{d\rho /dr} \nonumber\\
  i.e.,~~~  v_t^2-v_r^2 &=& \frac{dp_r}{d\rho}+\frac{d\Delta /dr}{d\rho /dr}
\end{eqnarray}
In order to maintain $-1\leq v_{t}^{2}-v_{r}^{2}\leq0$ throughout the fluid distribution it is required that $d\Delta/dr>0$ (from (\ref{crak})) as we have $d\rho/dr<0$ (see Fig. \ref{grad}),i.e., it is required that $\Delta$ is an increasing function of $r$ which is already satisfied by our model (see Fig. \ref{aniso}). With the help of graphical representation we have also shown that $v_t^2-v_r^2<0$ in Fig. \ref{stab} everywhere inside the fluid sphere which renders our model stable.

\begin{figure}[t] 
\begin{minipage}{\columnwidth}
\centering
\includegraphics[scale=0.6]{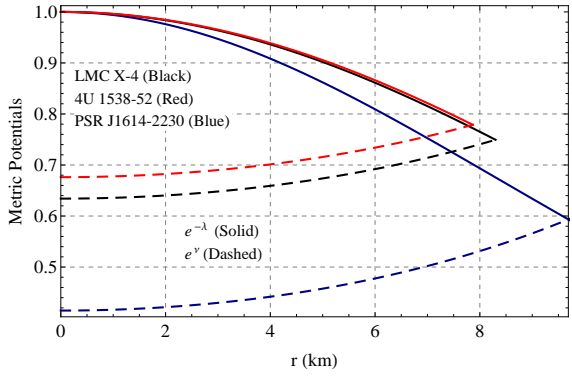}
\end{minipage}
\caption{Variation of metric potentials with radial coordinate $r$ for 4U1538-52, LMC X-4 and PSR J1614-2230 with their respective parameters given in Table \ref{tab}.}
\label{met}
\end{figure}

\begin{figure}[t] 
\begin{minipage}{\columnwidth}
\centering
\includegraphics[scale=0.6]{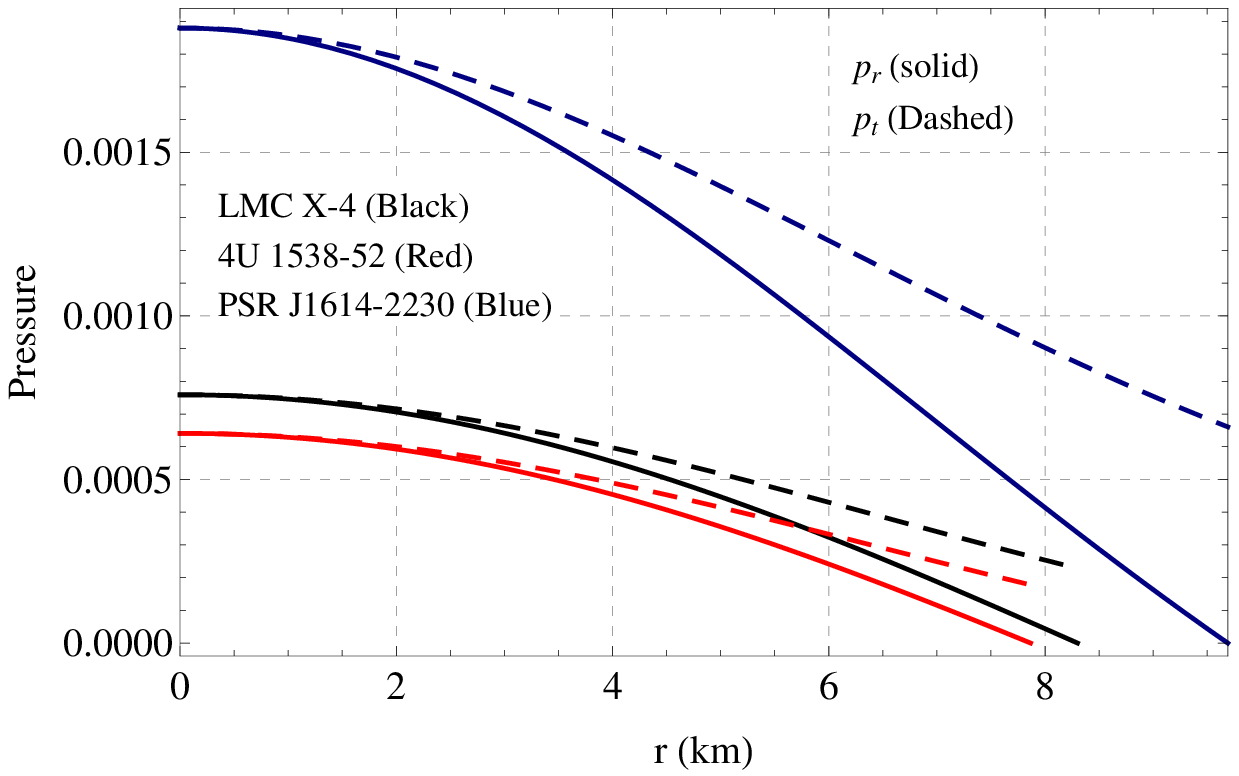}
\end{minipage}
\caption{Variation of interior pressures (km$^{-2}$) with radial coordinate $r$ for 4U1538-52, LMC X-4 and PSR J1614-2230 with their respective parameters given in Table \ref{tab}.}
\label{p}
\end{figure}

\begin{figure}[t] 
\begin{minipage}{\columnwidth}
\centering
\includegraphics[scale=0.6]{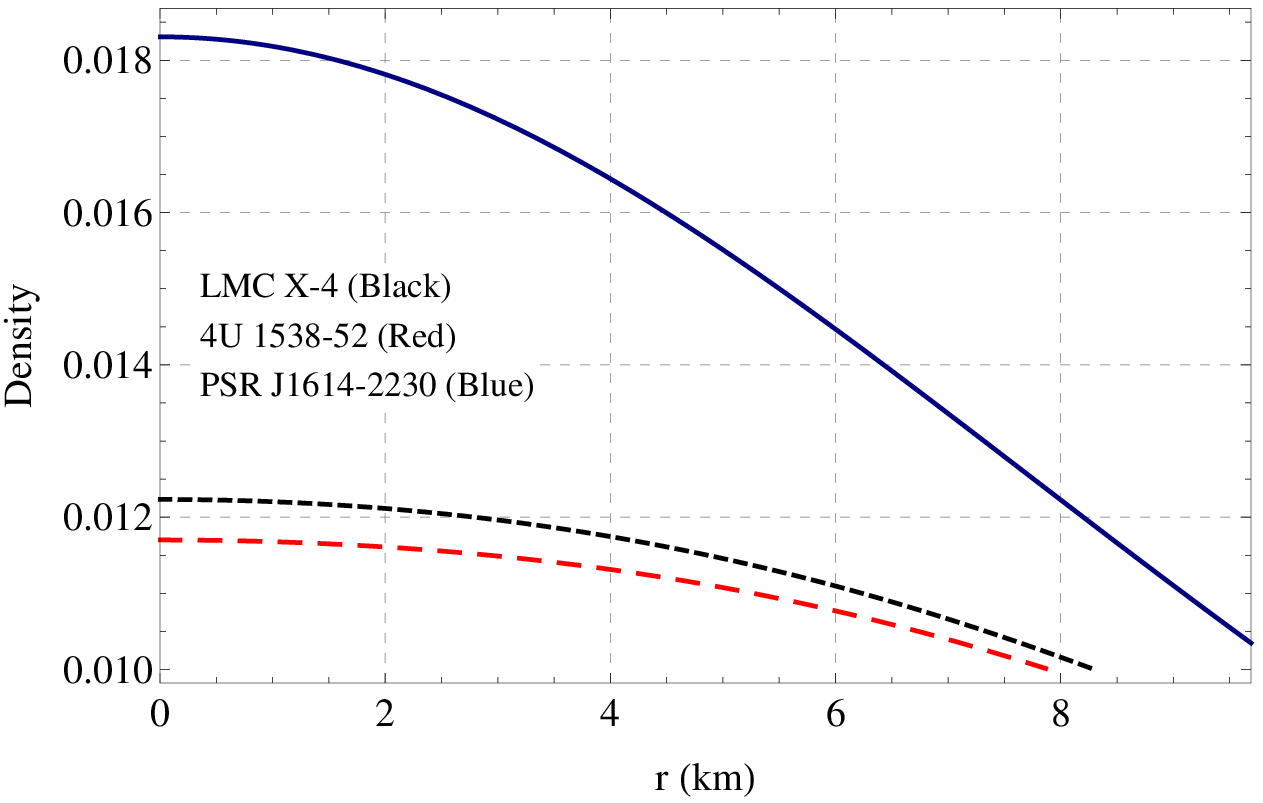}
\end{minipage}
\caption{Variation of density (km$^{-2}$)  with radial coordinate $r$ for 4U1538-52, LMC X-4 and PSR J1614-2230 with their respective parameters given in Table \ref{tab}.}
\label{rho}
\end{figure}

\begin{figure} 
\begin{minipage}{\columnwidth}
\centering
\includegraphics[scale=0.6]{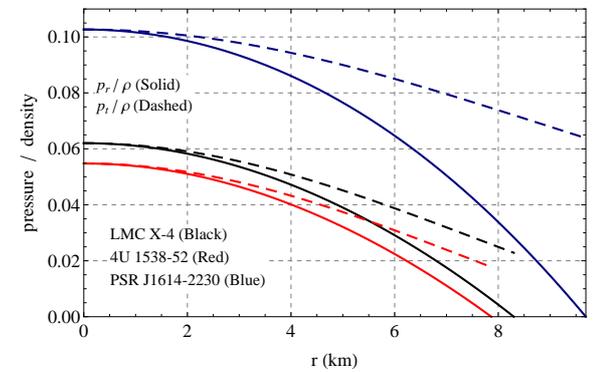}
\end{minipage}
\caption{Variation of pressure to density ratios  with radial coordinate $r$ for 4U1538-52, LMC X-4 and PSR J1614-2230 with their respective parameters given in Table \ref{tab}.}
\label{prho}
\end{figure}

\begin{figure} 
\begin{minipage}{\columnwidth}
\centering
\includegraphics[scale=0.6]{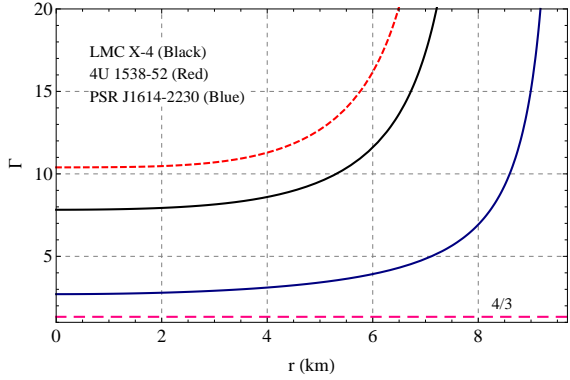}
\end{minipage}
\caption{Variation of relativistic adiabatic index  with radial coordinate $r$ for 4U1538-52, LMC X-4 and PSR J1614-2230 with their respective parameters given in Table \ref{tab}.}
\label{gamma}
\end{figure}

\begin{figure} 
\begin{minipage}{\columnwidth}
\centering
\includegraphics[scale=0.6]{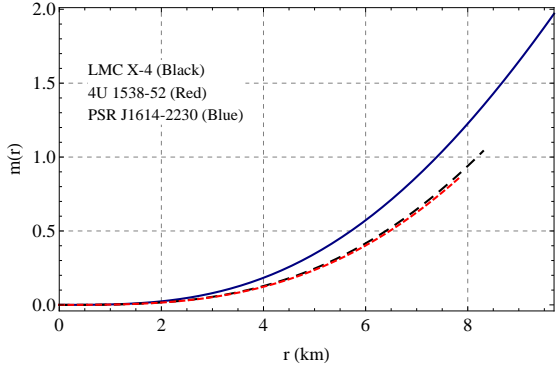}
\end{minipage}
\caption{Variation of interior mass with radial coordinate $r$ for 4U1538-52, LMC X-4 and PSR J1614-2230 with their respective parameters given in Table \ref{tab}.}
\label{mas}
\end{figure}

\begin{figure} 
\begin{minipage}{\columnwidth}
\centering
\includegraphics[scale=0.6]{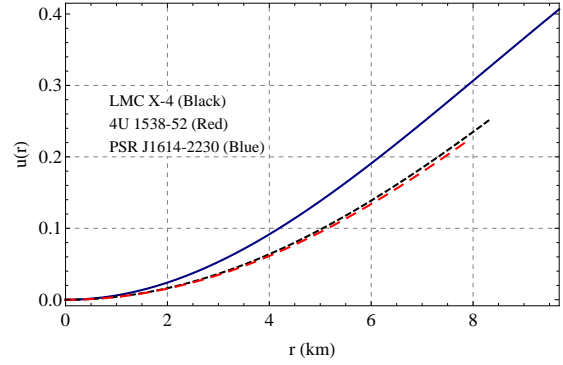}
\end{minipage}
\caption{Variation of compactness parameter with radial coordinate $r$ for 4U1538-52, LMC X-4 and PSR J1614-2230 with their respective parameters given in Table \ref{tab}.}
\label{com}
\end{figure}

\begin{figure} 
\begin{minipage}{\columnwidth}
\centering
\includegraphics[scale=0.6]{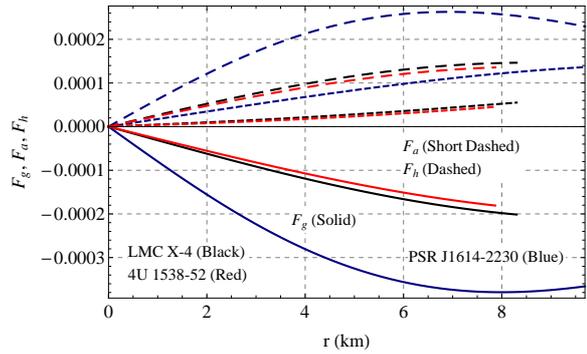}
\end{minipage}
\caption{Balancing of different forces in TOV equation for  static configurations of 4U1538-52, LMC X-4 and PSR J1614-2230 are plotted with radial coordinate $r$.}
\label{tov}
\end{figure}

\begin{figure} 
\begin{minipage}{\columnwidth}
\centering
\includegraphics[scale=0.6]{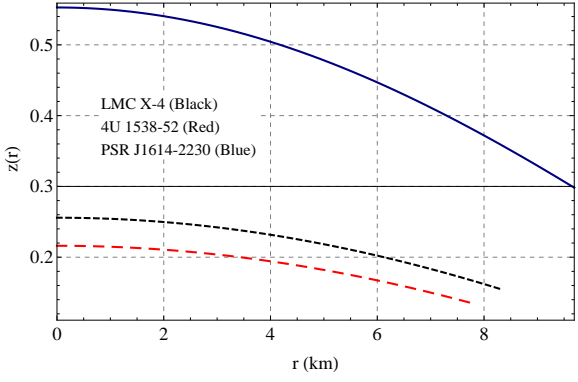}
\end{minipage}
\caption{Variation of red-shift with radial coordinate $r$ for 4U1538-52, LMC X-4 and PSR J1614-2230 with their respective parameters given in Table \ref{tab}.}
\label{red}
\end{figure}

\begin{figure} 
\begin{minipage}{\columnwidth}
\centering
\includegraphics[scale=0.6]{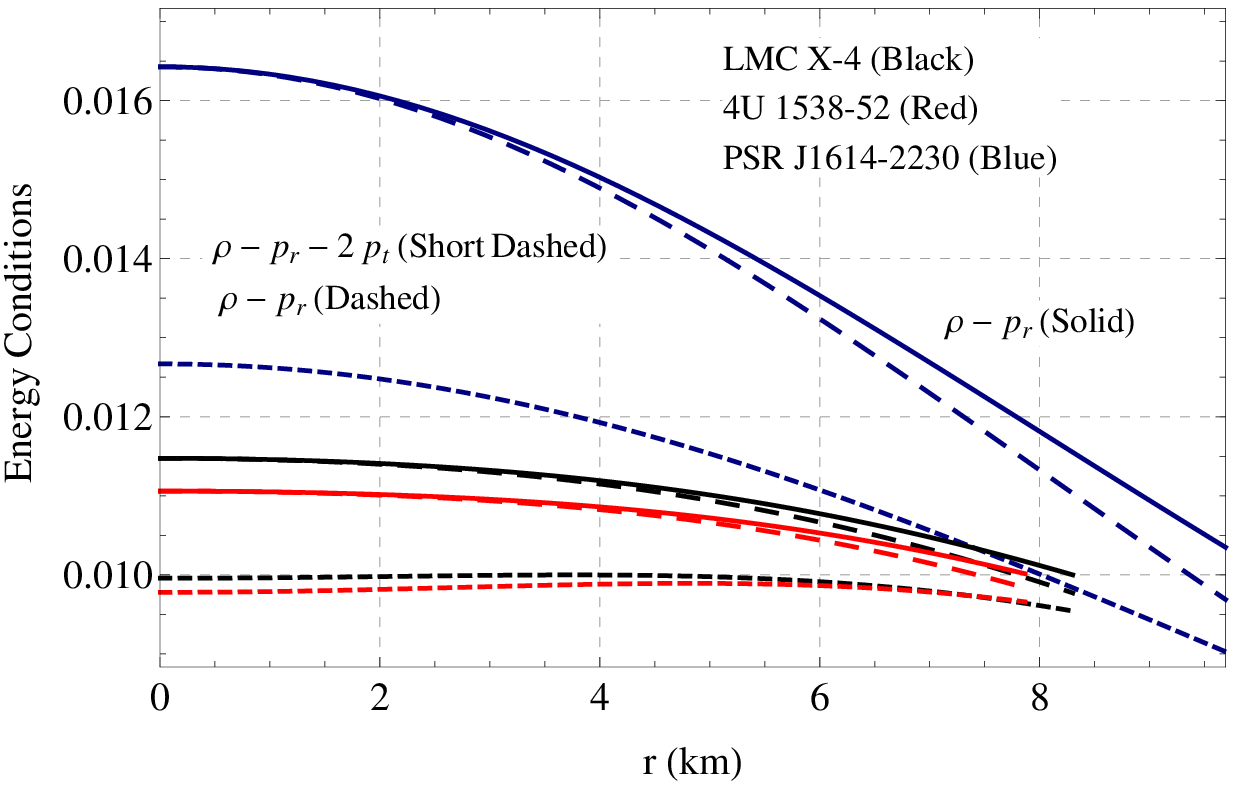}
\end{minipage}
\caption{Variation of $\rho-p_r,~\rho-p_t$ and $\rho-p-r-2p_t$ (km$^{-2}$) with radial coordinate $r$ for 4U1538-52, LMC X-4 and PSR J1614-2230 with their respective parameters given in Table \ref{tab}.}
\label{ec1}
\end{figure}

\begin{figure} 
\begin{minipage}{\columnwidth}
\centering
\includegraphics[scale=0.6]{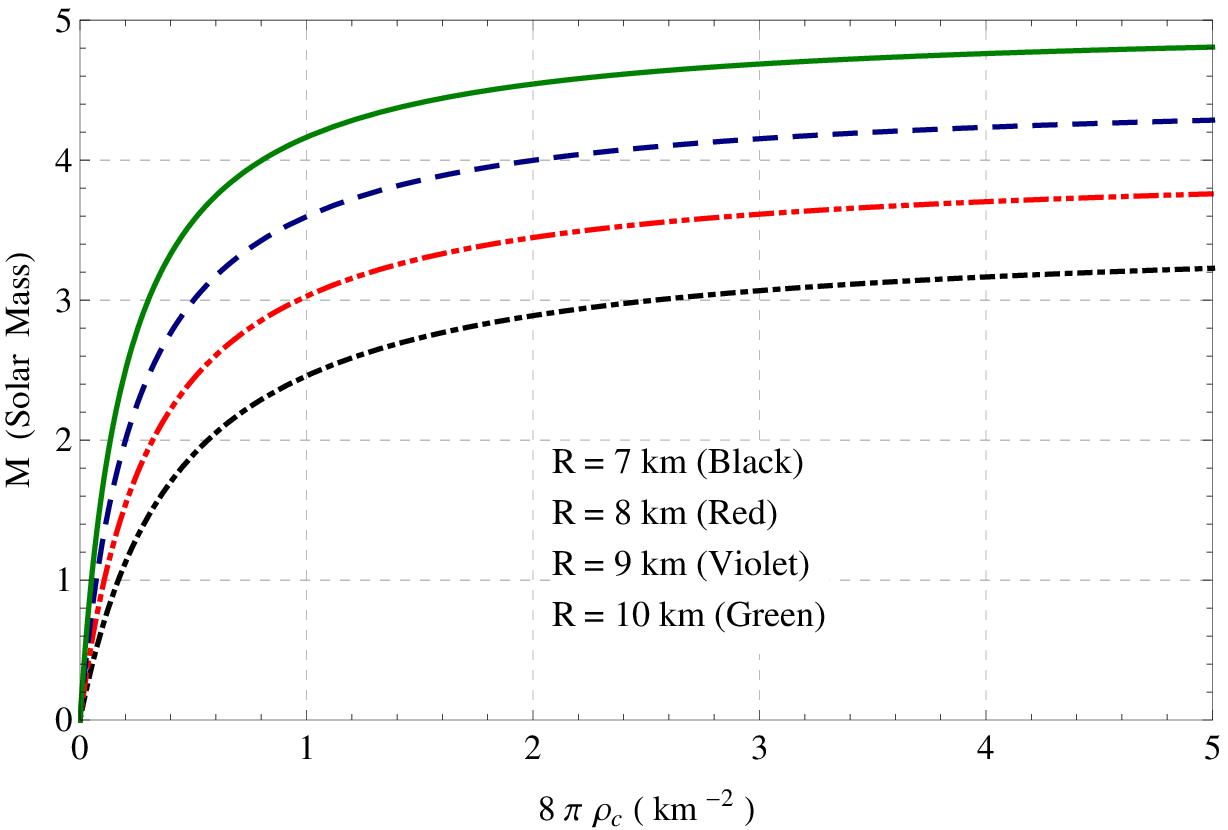}
\end{minipage}
\caption{Variation of mass with central density $8\pi \rho_c~(0-2.68 \times 10^{17}~g/cm^3)$ for $R=7-10~km$.}
\label{mrh}
\end{figure}

\subsection{Harrison-Zeldovich-Novikov static stability criterion}

The stability analysis adopted by \cite{chan64}, \cite{har65}, amongst other treatments requires the determination of eigen-frequencies of all the fundamental modes. However, \cite{har65} and \cite{zel} provide a simpler formalism to study the stability of the stellar model. They have assumed that the adiabatic index of a pulsating star is the same as in slowly deformed matter. This leads to a stable configuration only if the mass of the star is increasing with central density i.e. $dM/d\rho_c > 0$ and unstable if $dM/d\rho_c \le 0$.

In our model, the mass as a function of central density can be written as
\begin{eqnarray}
M = \frac{8\pi \rho_c R^3 \sin^2(c+bR^2)/3\sin^2c}{8\pi \rho_c R^2 /3\sin^2c -8\pi \rho_c R^2 \cos(2c+2bR^2)/3\sin^2c+2}\nonumber\\\label{mrho1}
\end{eqnarray}
which gives us (for a given radius)

\begin{equation}
{d M \over d \rho_c} = \frac{12 \pi  R^3 \sin ^2c ~~\sin ^2(b R^2+c)}{\left[8 \pi  R^2 \rho _c \sin ^2c ~~\sin ^2(b R^2+c)+3\right]^2} > 0.
\end{equation}

Fig. \ref{mrh} shows that our models are stable according to the static stability criterion. It is interesting to note that the stability of our configurations is enhanced with increasing radii and plateaus after attaining a maximum value for the respective central matter densities. Wherever the curve starts leveling off, implies that $dM/d\rho_c = 0$, indicating that the configuration is rendered unstable.

\section{Discussion of results}

Graphical analyses of the physical parameters $\big(e^{-\lambda},~p_r,$ $p_t,~\rho,$ $~p_r/\rho,~p_t/\rho,~v_r^2,~v_t^2,~Z\big)$ show that they are finite at the center and monotonically decreasing outward (Figs. \ref{met}, \ref{p}, \ref{rho}, \ref{prho}, \ref{sound}, \ref{red}). Figs. \ref{met}, \ref{aniso} and \ref{gamma}) show that $e^\nu$, anisotropy parameter, $\Delta$ and $\Gamma$ are increasing radially outward.

The Null Energy Condition $\big(\rho-p_i \ge 0\big)$, Dominant Energy Condition $\big(\rho-p_i \ge 0$, $\rho \ge 0\big)$ and Strong Energy Condition $\big(\rho-p_i \ge 0$, $\rho-p_r-2p_t \ge 0\big)$ are simultaneously satisfied by our solution (Fig. \ref{ec1}). The solution can also represent static and stable stellar configurations as the stability factor $v_t^2-v_r^2$ lies between the limits $-1$ to 0, (Fig. \ref{stab}). For a non-collapsing stellar configuration, the adiabatic index must also be greater than 4/3 for positive values of anisotropy which can be seen from Fig. \ref{gamma}. Furthermore, the gravitational force $F_g$ in the configuration is balanced by the combined effect of hydrostatic $F_h$ and anisotropic $F_a$ force (Fig. \ref{tov}) and thus the solution satisfies the TOV-equation Eq. (\ref{tove}). The mass and the compactness parameter are also monotonically increase from the center to the surface of the star and the compactness parameter is also within the Buchdahl limit i.e. $u \le 8/9$ (Figs. \ref{mas}, \ref{com}). The negative values of the gradients of density and pressures signify that the density and pressures are decreasing radially outward (Fig. \ref{grad}).

The well-behaved nature of the solution depends on the parameter $c$ for a particular star. For 4U 1538-52 the solution behaves well for $0.1574 \le c \le 0.46$ and for the values of $a,~b,~A,~B,~M,~R$ given in Table \ref{tab}, which corresponds to $1\ge v_{r0}^2 \ge 0.13$, $0.91 \ge v_{t0}^2 \ge 0.04$ and $17.8 \ge \Gamma_0 \ge 3.8$. For LMC X-4 the solution behaves well for $0.1235 \le c \le 0.35$ and for the values of $a,~b,~A,~B,~M,~R$ given in Table \ref{tab}, that yields to $0.99\ge v_{r0}^2 \ge 0.15$, $0.91 \ge v_{t0}^2 \ge 0.06$ and $15.35 \ge \Gamma_0 \ge 3.35$. And finally for PSR J1614- 2230 the solution behaves well for $0.05 \le c \le 0.13$ along with the values of $a,~b,~A,~B,~M,~R$ given in Table \ref{tab}, corresponds to $1\ge v_{r0}^2 \ge 0.21$, $0.94 \ge v_{t0}^2 \ge 0.13$ and $8.34 \ge \Gamma_0 \ge 2.34$. Hence, we can conclude that smaller values of $c$ leads to stiffer equation of states and vice-versa. The calculated masses and radii of the present stars are well fitted with those provided by Gangopadhyay et al. \cite{gango}.

\begin{acknowledgements}
Authors are grateful to the anonymous referee(s) for rigorous review, constructive comments and useful suggestions. First two authors also acknowledge their gratitude to Air Marshal J. S. Kler, VM, Commandant, NDA, for his motivation and encouragement.
\end{acknowledgements}

\end{document}